\documentclass{iau}
\usepackage{graphicx}

\newcommand{\Msolar}{\mbox{\,$\rm M_{\odot}$} }	      % solar mass
\newcommand{\subs}[1]{$_{\rm #1}$}
\newcommand{\sups}[1]{$^{\rm #1}$}

\title[A Bcool spectropolarimetric survey of over 150 solar-type stars] %% give here short title %%
{A Bcool spectropolarimetric survey of over 150 solar-type stars}

\author[Marsden et al.]   %% give here short author list %%
{Stephen Marsden,$^1$ Pascal Petit,$^{2,3}$ Sandra Jeffers,$^4$\\
Jose-Dias do Nascimento,$^5$ Bradley Carter$^1$ \and Carolyn Brown$^1$\\
on behalf of the Bcool project team}

\affiliation{$^{1}$Computational Engineering and Science Research Centre, University of Southern Queensland, Toowoomba, 4350, Australia\\ email: {\tt Stephen.Marsden@usq.edu.au}\\
$^{2}$Universit\'{e} de Toulouse, UPS-OMP, Institut de Recherche en Astrophysique et Plan\'{e}tologie, Toulouse, France\\
$^{3}$CNRS, Institut de Recherche en Astrophysique et Plan\'{e}tologie, 14 Avenue Edouard Belin, F-31400 Toulouse, France\\
$^{4}$Institut f\"{u}r Astrophysik, Georg-August-Universit\"{a}t G\"{o}ttingen, Friedrich-Hund-Platz 1, 37077 G\"{o}ttingen, Germany\\
$^{5}$Departamento de Fisica Te\'{o}rica e Experimental, Universidade Federal do Rio Grande do Norte, CEP: 59072-970 Natal, RN, Brazil}

\pubyear{2013}
\volume{302}  %% insert here IAU Symposium No.
\pagerange{119--126}
\jname{Magnetic Fields Throughout Stellar Evolution}
\editors{P. Petit, M.M. Jardine \& H.C. Spruit, eds.}
\begin{document}

\maketitle

\begin{abstract}
As part of the Bcool project, over 150 solar-type stars chosen mainly from planet search databases have been observed between 2006 and 2013 using the NARVAL and ESPaDOnS spectropolarimeters on the Telescope Bernard Lyot (Pic du Midi, France) and the Canada France Hawaii Telescope (Mauna Kea, USA), respectively. These single ``snapshot'' observations have been used to detect the presence of magnetic fields on 40\% of our sample, with the highest detection rates occurring for the youngest stars. From our observations we have determined the mean surface longitudinal field (or an upper limit for stars without detections) and the chromospheric surface fluxes, and find that the upper envelope of the absolute value of the mean surface longitudinal field is directly correlated to the chromospheric emission from the star and increases with rotation rate and decreases with age.
%% add here a maximum of 10 keywords, to be taken form the file <Keywords.txt>
\end{abstract}

\firstsection % if your document starts with a section,
              % remove some space above using this command.
\section{Introduction}

\begin{figure}[t]
% \vspace*{-2.0 cm}
\begin{center}
 \includegraphics[width=6.7cm]{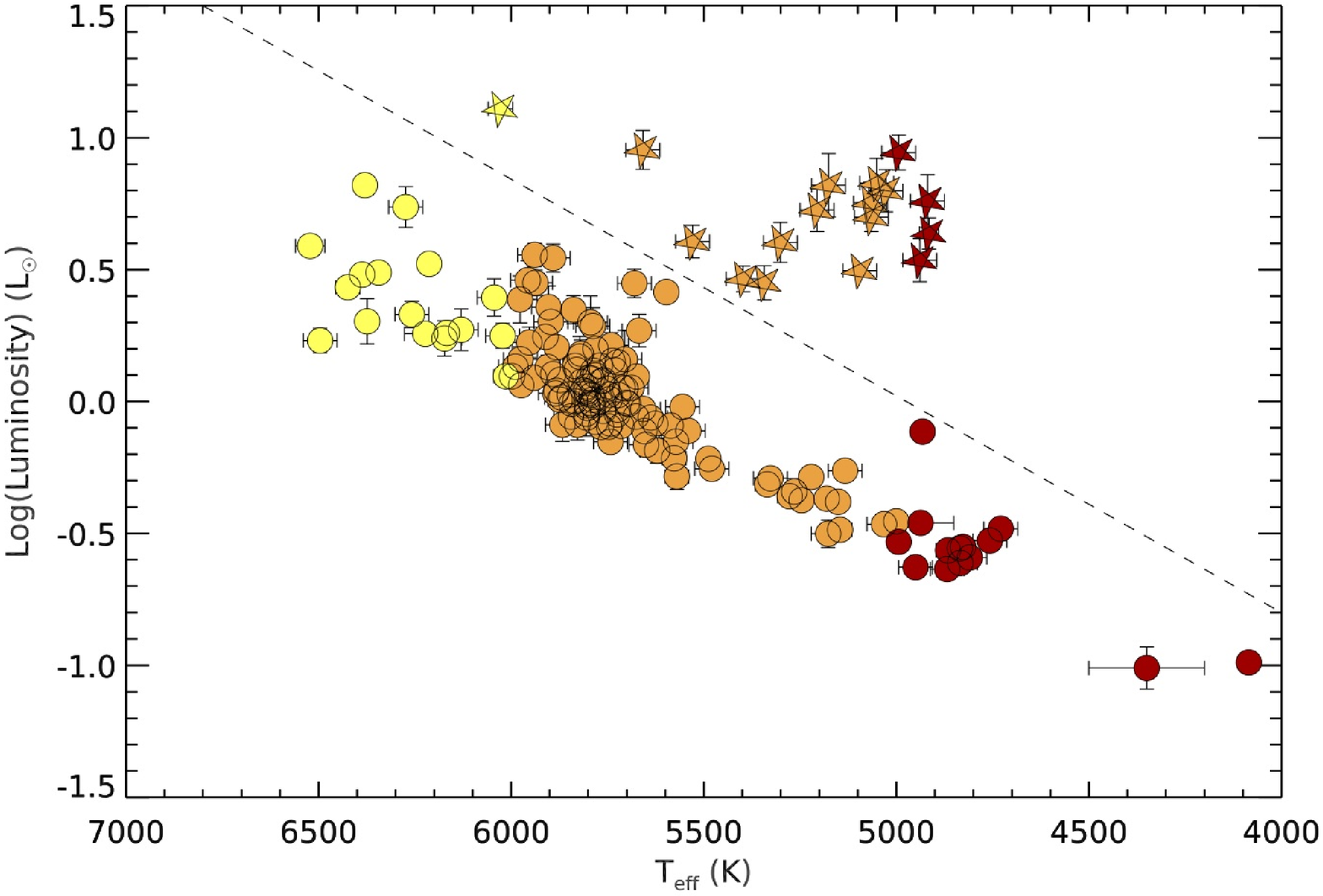}
 \includegraphics[width=6.7cm]{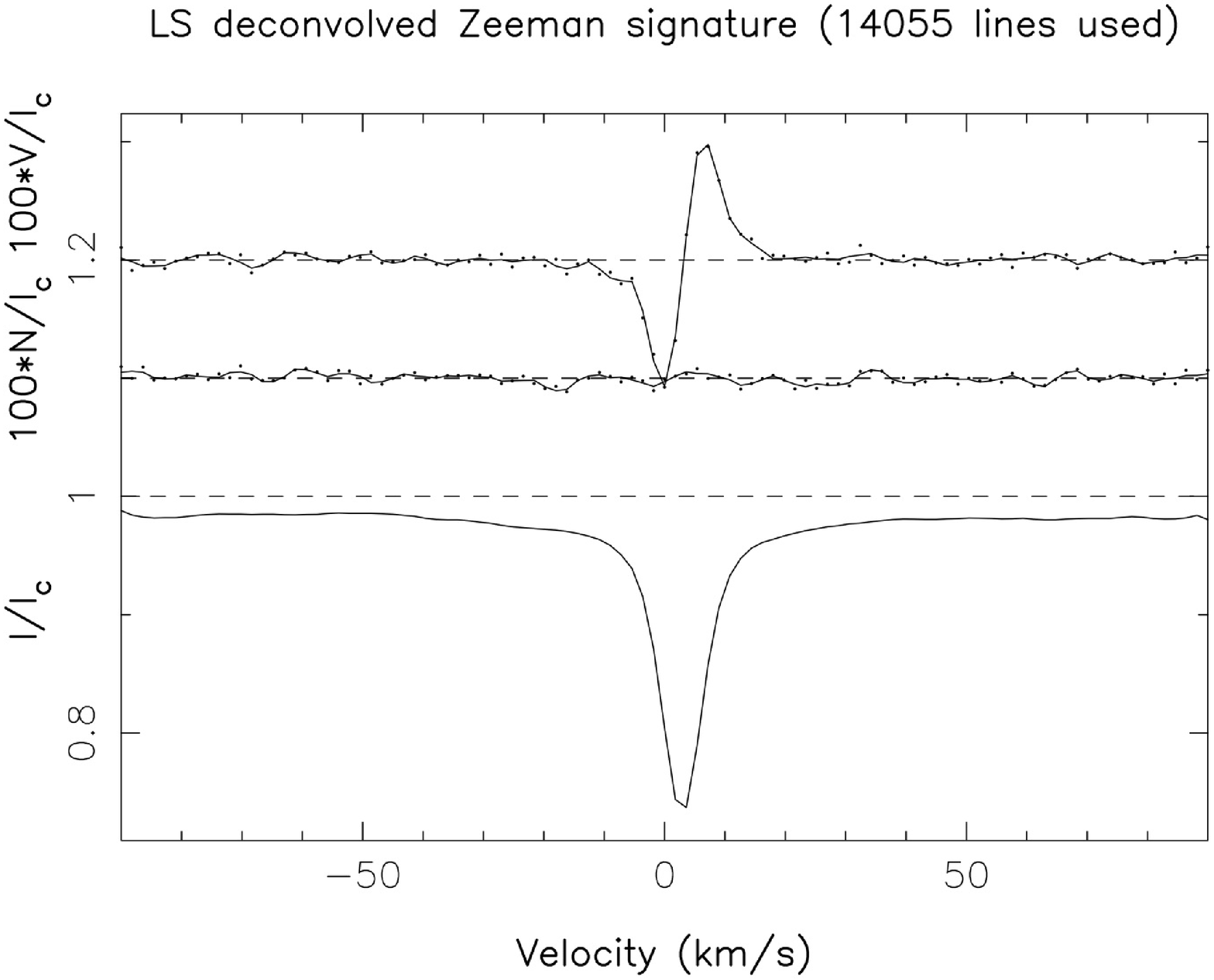}
% \vspace*{-1.0 cm}
 \caption{The left-hand plot shows the HR diagram for our stars, with stars in red (dark grey) having T\subs{eff} $<$ 5000 K, those in orange (mid grey) having 5000 K $\le$ T\subs{eff} $\le$ 6000 K and those in yellow (light grey) having T\subs{eff} $>$ 6000 K. Filled circles represent dwarf stars while five-pointed stars represent subgiants, with the dashed line showing the dividing line between the two. The right-hand plot shows the Stokes V (upper), Null (middle) and Stokes I (lower) LSD profiles of a sample star with the Stokes V profile showing a magnetic detection.}
   \label{fig1}
\end{center}
\end{figure}

The Bcool\footnote{http://bcool.ast.obs-mip.fr} project is an international collaboration looking at the magnetic activity of low-mass stars, predominately through the use of spectropolarimetry. One of the main research areas of Bcool is the study of solar-type stars to help understand how the magnetic dynamo operates in such stars. To this end the Bcool project has obtained high-resolution spectropolarimetric observations of 167 solar-type stars using the twin spectropolarimeters NARVAL and ESPaDOnS on the Telescope Bernard Lyot and the Canada France Hawaii Telescope, respectively. The observations were taken over 25 observing runs from late 2006 through to mid 2013. The targets are predominately dwarf stars ($\sim$90\%) although a number (17) were identified to be subgiants, see Figure~\ref{fig1} (left-hand side). The sample covers stellar masses from 0.6 up to 2.5 \Msolar with a range of ages. A significant fraction of the sample (44\%) have stellar effective temperatures around the solar value (i.e. 5750 $\pm$ 125 K). The main aims of this survey are to determine which stars have detectable magnetic fields for follow-up study through the use of Zeeman Doppler Imaging (ZDI, \cite[Donati et al. 1997]{DonatiJF:1997}) to map their surface magnetic topologies and also to determine if the large-scale magnetic field properties of solar-type stars vary with basic stellar parameters. This paper describes the preliminary results from this spectropolarimetric solar-type star ``snapshot'' survey with the main results to be presented in a later paper (Marsden et al. in prep.).

\section{Targets and Observations}

The targets in the sample have been primarily taken from the Spectroscopic Properties of Cool Stars (SPOCS) sample (\cite[Valenti \& Fischer 2005]{ValentiJA:2005}, \cite[Takeda et al. 2007]{TakedaG:2007}) with additional stars taken from \cite[Wright et al. (2004)]{WrightJT:2004} and \cite[Baliunas et al. (1995)]{BaliunasSL:1995}.

Each star was observed at least once using the NARVAL and ESPaDOnS spectropolarimeters and from each observation both an intensity (Stokes I) and a circularly polarsied (Stokes V) spectrum was recovered using {\sc Libre-Esprit,} an automatic reduction software package based on {\sc Esprit} (\cite[Donati et al. 1997]{DonatiJF:1997}). As the signal-to-noise ratio of the observed spectra was not high enough to detect Zeeman signatures in individual lines it was necessary to apply the technique of Least-Squares Deconvolution (LSD, \cite[Donati et al. 1997]{DonatiJF:1997}) to the polarimetric signals from the data. A magnetic field can then be detected through the variations in the Stokes V LSD profile, see Figure~\ref{fig1} (right-hand side).

\begin{figure}[t]
% \vspace*{-2.0 cm}
\begin{center}
 \includegraphics[width=6.7cm]{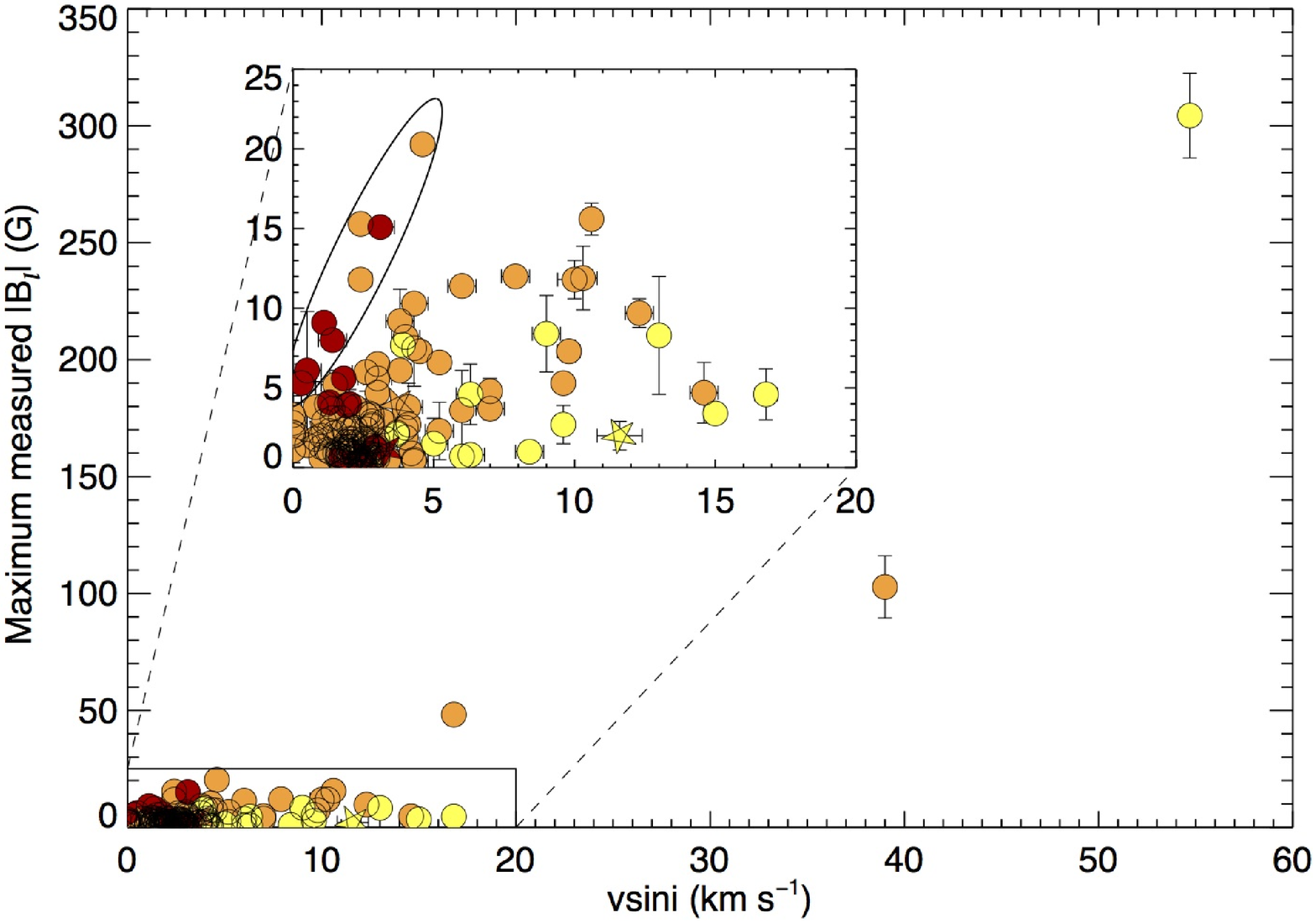}
 \includegraphics[width=6.7cm]{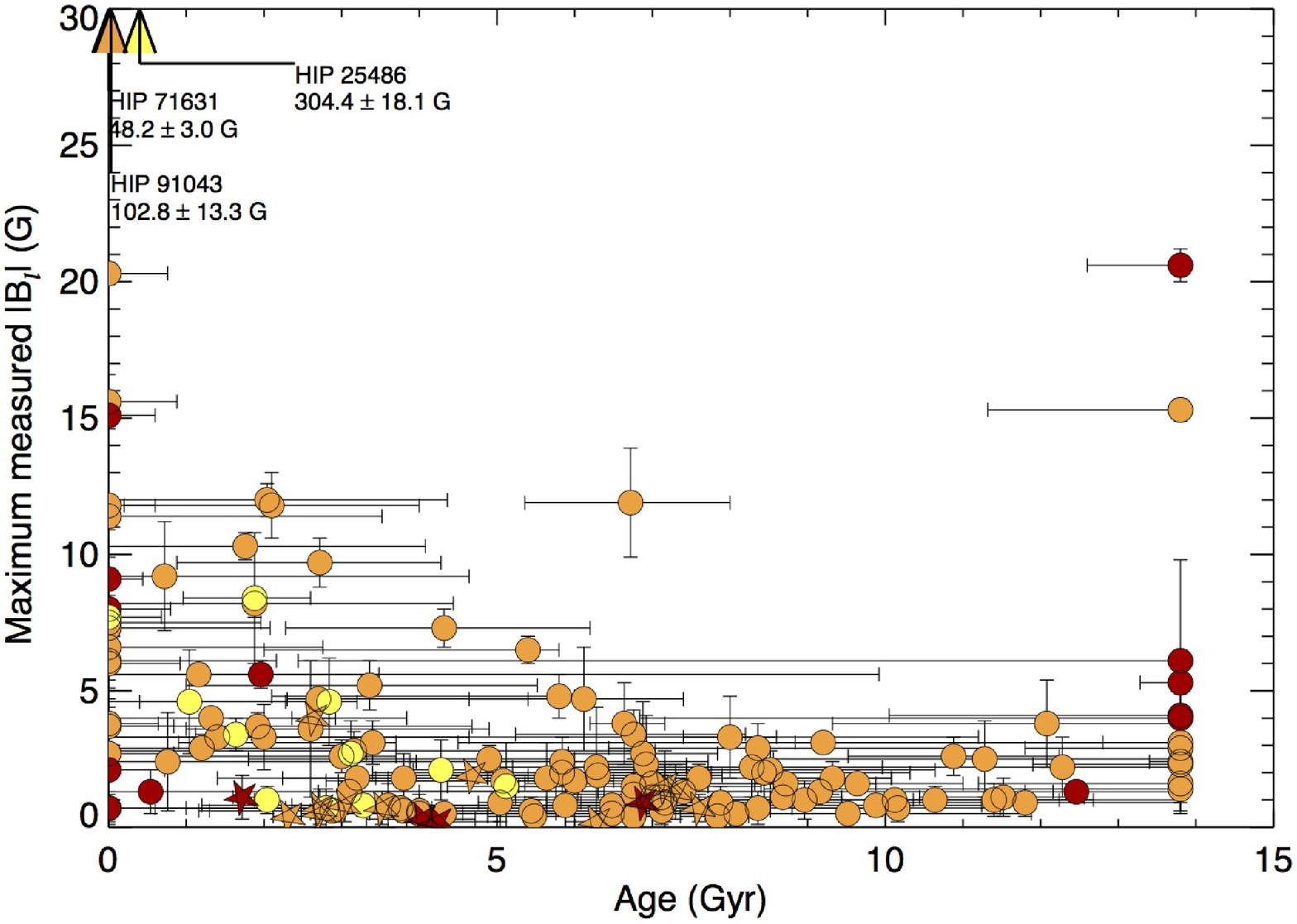}
% \vspace*{-1.0 cm}
 \caption{Plot of the absolute value of the mean longitudinal magnetic field ($|B_{l}|$) against $\!${\em v\,}sin{\em i} (left-hand side) and stellar age (right-hand side). The symbols are the same as in Figure~\ref{fig1} (left-hand side).}
   \label{fig2}
\end{center}
\end{figure}

\section{Results}

For our sample of 167 solar-type stars, we have been able to detect a magnetic field on the surface of 40\% of our targets (67 stars). This includes the detection of magnetic fields on 3 of our 17 subgiant stars. There appears to be a slight increase in the detection rate for K-stars (12/21 = 57\%) over G-stars (48/128 = 38\%) and F-stars (8/18 = 44\%), but the numbers of K- and F-stars in our sample are small compared to the G-stars. Those stars with magnetic detections have a wide range of rotation rates ($\!${\em v\,}sin{\em i} = 0.0 to 54.7 km s\sups{-1}) and cover all ages, with 21 of our stars with magnetic field detections being classified as mature solar-type stars with ages greater than 2 Gyr.  

\subsection{The Longitudinal Magnetic Field}

A measure of the strength of the surface magnetic field on a star can be derived from the star's Stokes V and I LSD profiles using (\cite[Donati et al. 1997]{DonatiJF:1997}):
\begin{equation}
B_{l} = -2.14 \times 10^{11} \frac{\int vV(v)dv}{\lambda g c \int [I_{c} - I(v)]dv}, \label{eqn1}
\end{equation}
where $B_{l}$ is the line-of-sight-component of the stellar magnetic field integrated over the visible stellar disc, $c$ is the speed of light and $\lambda$ and $g$ are the mean wavelength and Land\'{e} factor of the LSD profile. $I_{c}$ is the continuum level of the intensity profile and both the Stokes V and I LSD profiles are integrated over velocity space ($v$).

The upper envelope of the absolute value of the longitudinal magnetic field ($|B_{l}|$) is found to increase with rotation rate and decrease with age (see Figure~\ref{fig2}), as would be expected of an indicator of the surface magnetic activity. However, unlike traditional activity indicators which are sensitive to the magnetic field strength, $|B_{l}|$ is also sensitive to the polarity of the magnetic field and it thus very dependent upon the distribution and polarity mix of magnetic regions across the stellar surface. Therefore, due to cancellation effects, the measure of $|B_{l}|$ can be lower than expected. This explains why it is the upper envelope of $|B_{l}|$ that correlates with stellar parameters.

\subsection{Calcium HK emission}

In addition to the longitudinal magnetic field we have also analysed more traditional activity measures, including the Calcium HK emission from the star. The Ca HK emission was standardised to the Mt. Wilson survey's S-index (cf., \cite[Baliunas et al. 1995]{BaliunasSL:1995}) using the method described in \cite[Wright et al. (2004)]{WrightJT:2004}, by calibrating our results to those of \cite[Wright et al. (2004)]{WrightJT:2004} using common stars between the two samples.  

\begin{figure}[t]
% \vspace*{-2.0 cm}
\begin{center}
 \includegraphics[width=6.7cm]{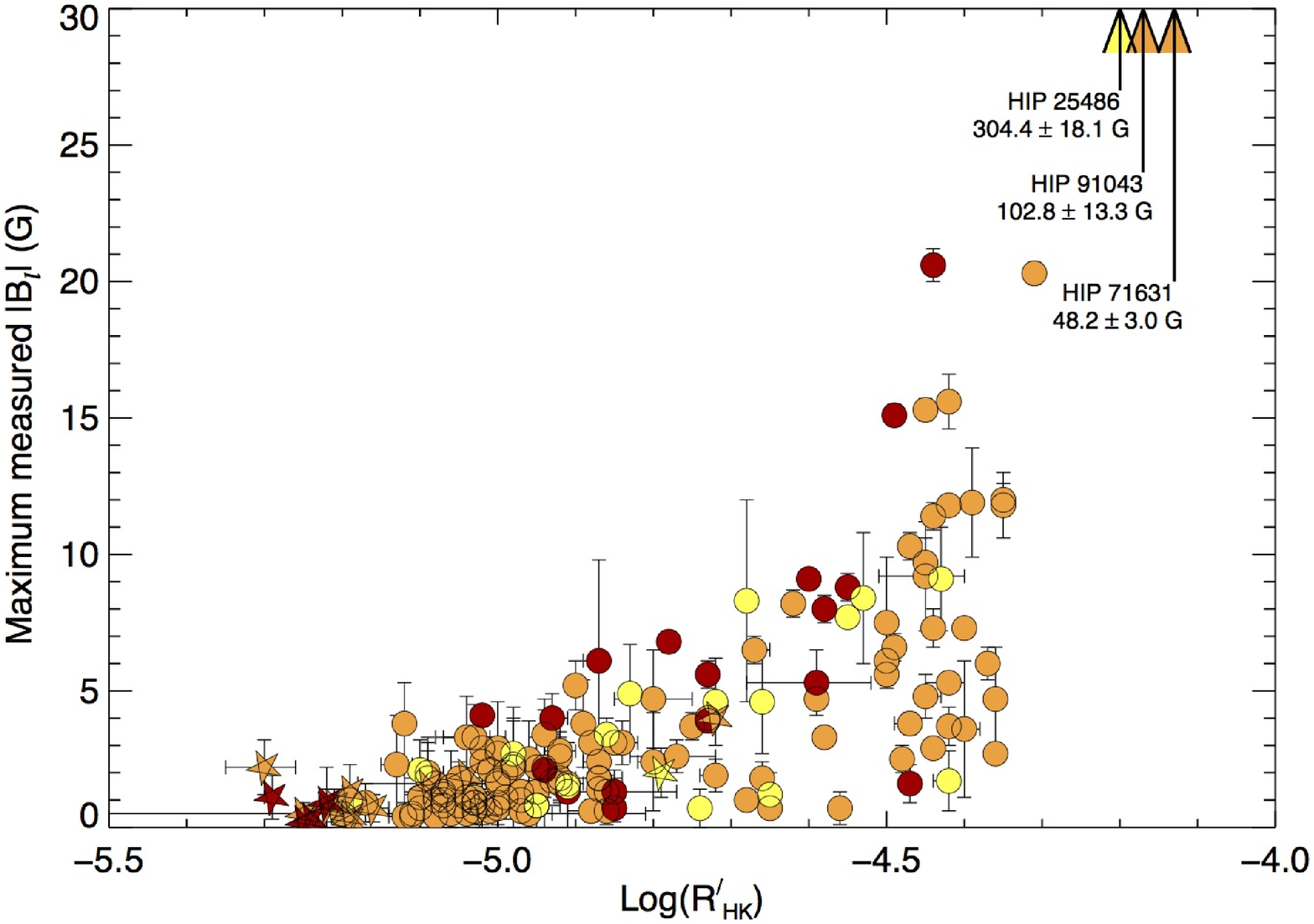}
 \includegraphics[width=6.7cm]{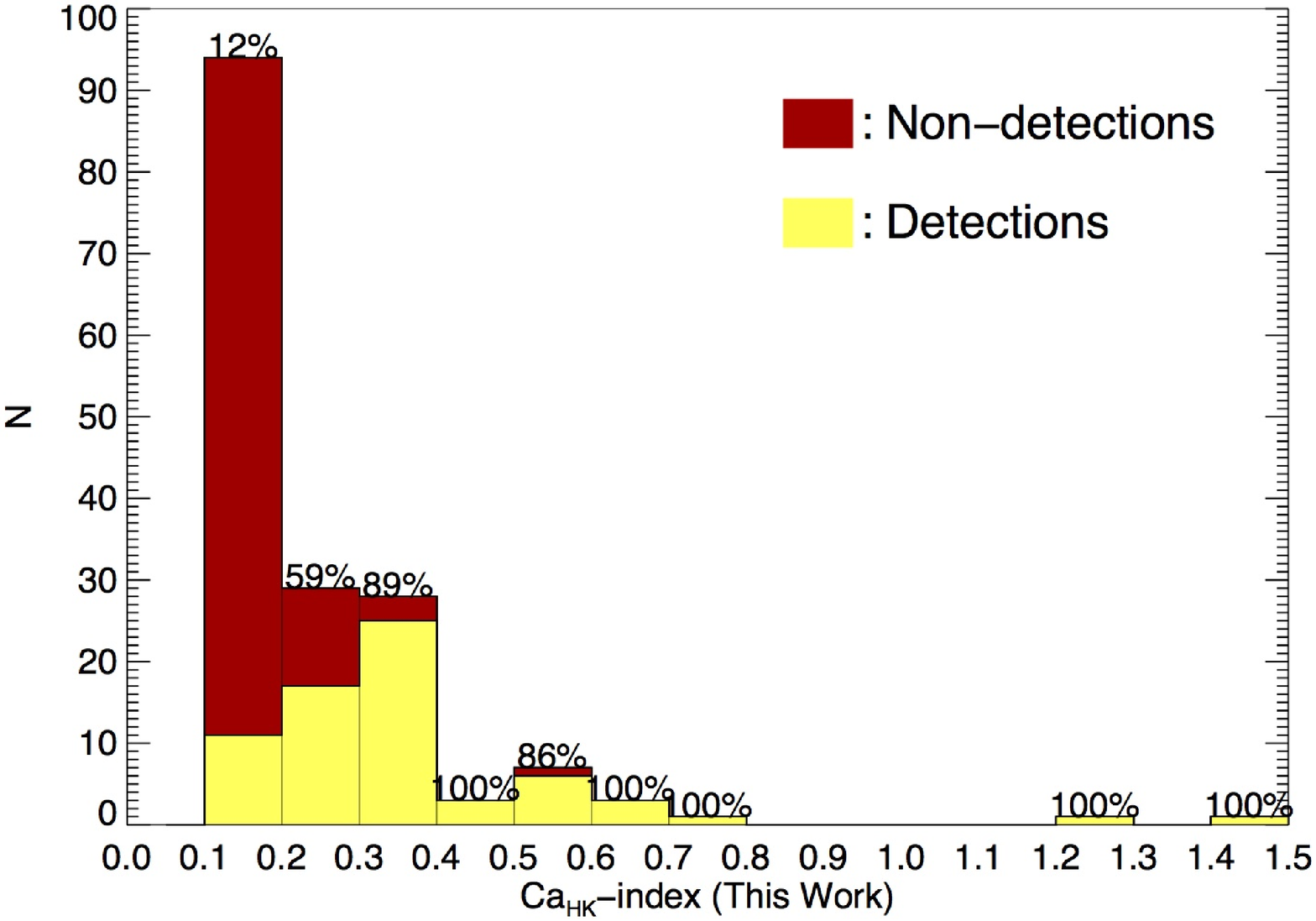} 
% \vspace*{-1.0 cm}
 \caption{Plot of $|B_{l}|$ against Ca HK emission (left-hand side) and a Histogram of magnetic field detections/non-detections against Ca HK emission (right-hand side). The symbols in the left-hand side plot are the same as in Figure~\ref{fig1} (left-hand side), while the percentages above each column in the histogram gives the detection rate for each bin.}
   \label{fig3}
\end{center}
\end{figure}

Since $B_{l}$ is based on an average measure of the amount of magnetic flux on the visible stellar surface it is expected that $B_{l}$ should correlate with other activity indicators, such as the Ca HK emission. $|B_{l}|$ is plotted against the simultaneously obtained Ca HK emission for our stars in Figure~\ref{fig3} (left-hand side) with the plot showing that there is a clear correlation between the upper envelope of $|B_{l}|$ and the Ca HK emission.

The right-hand side plot in Figure~\ref{fig3} shows a histogram of the magnetic field detections / non-detections against the Ca HK-index. As can be seen the detection rate increases with the Ca HK-index and for stars with a Ca HK-index greater than 0.3 the detection rate is almost 90\%, while for those stars with Ca HK-index values less than 0.2 the detection rate is down near 10\%.

\section{Conclusions}

One of the most active research areas in the Bcool project is the study of solar-type stars. We have observed a large sample (167) of such stars using high-resolution spectropolarimetric observations  in order to detect and characterise their magnetic fields. Prior to this survey, the majority of the magnetic field detections on solar-type stars have been for young stars (i.e. \cite[Donati et al. 1997]{DonatiJF:1997}, \cite[Donati et al. 2003]{DonatiJF:2003}, \cite[Marsden et al. 2006]{MarsdenSC:2006}, \cite[Marsden et al. 2011]{MarsdenSC:2011}). This project has detected magnetic fields on 40\% of our sample (67 stars), with 21 of our detected stars being classified as mature age solar-type stars with an age greater than 2 Gyr. This is a four-fold increase in the number of mature-age stars with magnetic field detections discovered so far. In addition, we have detected magnetic fields on 3 of our 17 subgiant stars.

We have shown that the upper envelope of the absolute value of the longitudinal magnetic field ($|B_{l}|$) increases with rotation rate and decreases with age and is strongly correlated to other more traditional activity indicators. We have also shown that the detection rate for magnetic fields is strongly linked to the Ca HK emission of the star.

This survey represents a unique dataset on the magnetic fields of solar-type stars that will provide the basis for further detailed study into the magnetic dynamos of these stars.


\begin{thebibliography}{}
\bibitem[Baliunas \etal (1995)]{BaliunasSL:1995} {Baliunas S. L., Donahue R. A., Soon W. H., et al.} 1995, ApJ, 438, 269
\bibitem[Donati \etal (1997)]{DonatiJF:1997} {Donati J.-F., Semel M., Carter B. D., Rees D. E., \& Cameron A. C.} 1997, MNRAS, 291, 658
\bibitem[Donati \etal (2003)]{DonatiJF:2003} {Donati J.-F., Collier Cameron A., Semel M., et al.} 2003, MNRAS, 345, 1145
\bibitem[Marsden \etal (2006)]{MarsdenSC:2006} {Marsden S. C., Donati J.-F., Semel M., Petit P., Carter B. D.} 2006, MNRAS, 370, 468 
\bibitem[Marsden \etal (2011)]{MarsdenSC:2011} {Marsden S. C., Jardine M. M., Ram\'{i}rez V\'{e}lez J. C., et al.} 2011, MNRAS, 413, 1922
\bibitem[Takeda \etal (2007)]{TakedaG:2007} {Takeda G., Ford E. B., Sills A., Rasio F. A., Fischer D. A., \& Valenti J. A.} 2007, ApJS, 168, 297
\bibitem[Valenti \& Fischer (2005)]{ValentiJA:2005} {Valenti J. A., \& Fischer D. A.} 2005, ApJS, 159, 141
\bibitem[Wright \etal (2004)]{WrightJT:2004} {Wright J. T., Marcy G. W., Butler R. P., \& Vogt S. S.} 2004, ApJS, 152, 261
\end{thebibliography}
\end{document}